%% file: PRD-final.tex
\documentclass[aps,prd,twocolumn,floatfix,nofootinbib,showpacs,superscriptaddress]{revtex4-1}

\usepackage[makeroom]{cancel} 

\usepackage{amsmath}
\usepackage{amssymb}
\usepackage{amsfonts}
\usepackage{graphicx}
\usepackage{color}
\usepackage{dsfont}
\usepackage[normalem]{ulem}
\usepackage{hyperref}
\usepackage{mathtools}
\usepackage{mathrsfs}
\usepackage{comment}
\usepackage[utf8]{inputenc}
\usepackage{multirow} 
\usepackage{subfigure}
\usepackage{varioref}
\usepackage[active]{srcltx}

\expandafter\ifx\csname package@font\endcsname\relax\else
\expandafter\expandafter
\expandafter\usepackage
\expandafter\expandafter
\expandafter{\csname package@font\endcsname}%
\fi

\hypersetup{
	colorlinks=true,       
	linkcolor=blue,          
	citecolor=magenta,        
	filecolor=cyan,      
	urlcolor=magenta           
}

\usepackage[section]{placeins}

\usepackage{fancybox}
\labelformat{figure}{Fig.~#1}

\begin{document}
\title{
Probing Axion–Photon conversion via circular polarization imprints in \\the CMB $V$-mode observations} 
\author{Ashu Kushwaha} 
\email{kushwaha.a.ce1b@m.isct.ac.jp}
\affiliation{Department of Physics, Institute of Science Tokyo, 2-12-1 Ookayama, Meguro-ku,
Tokyo 152-8551, Japan}
\author{Rajeev Kumar Jain}
\email{rkjain@iisc.ac.in}
\affiliation{Department of Physics, Indian Institute of Science C. V. Raman Road, Bangalore 560012, India}
%
	%
	%
\begin{abstract}

In the presence of a background magnetic field, axions or axion-like particles (ALPs) can be resonantly converted to photons when their mass is nearly equal to the effective photon mass. In this paper, we propose a novel method to constrain the parameter space of ALPs by investigating the resulting imprints of axion-photon conversion in the cosmic microwave background (CMB) observations. We show that a helical magnetic field existing prior to the CMB epoch can generate an excess population of photons carrying net circular polarization due to the axion-photon conversion mechanism.
Consequently, current measurements of the angular power spectrum of circular polarization ($V$-mode) in the CMB can be used to constrain the parameter space of ALP mass and its coupling to photons.
In the optimistic scenario of a maximally helical magnetic field with strength $\sim {\rm nG}$, we find that CLASS observations at $40 \, {\rm GHz}$ provide the most stringent upper limits to date, on the previously unconstrained regions of axion-photon coupling corresponding to ALP masses in the range $10^{-10}-10^{-8} \, {\rm eV}$.

\end{abstract}
\maketitle

\section{Introduction}
Precision measurements of the cosmic microwave background (CMB) radiation have been instrumental in 
establishing the $\Lambda$ Cold Dark Matter ($\Lambda$CDM) framework as the concordance model of the Universe. The CMB temperature and polarization (linear and circular) anisotropies provide crucial insights into the  origin, evolution, composition and the fundamental physical processes
of the Universe~\cite{Kosowsky:1994cy,Hu:1997hv,Kamionkowski:2015yta,Book-Durrer-2020,Komatsu:2022nvu}. For instance, B-mode polarization can probe the primordial gravitational waves (GW) produced during inflation and constrain the inflationary energy scale while a nonzero E-B correlation would signal parity-violating processes in the early Universe~\cite{Kosowsky:1994cy,Hu:1997hv,Kamionkowski:2015yta,Komatsu:2022nvu}. Thomson scattering at recombination generates only linear polarization (characterized by the Stokes $Q$ and $U$ parameters), implying vanishing primordial circular polarization (i.e., Stokes $V$ parameter).
However, $V$-modes can arise from various mechanisms in the early Universe, such as photon-photon scattering~\cite{Sadegh:2017rnr,2019-Inomata.Kamionkowski-PRL,Hoseinpour:2020hic}, photon-graviton scattering~\cite{Bartolo:2018igk}, Faraday conversion~\cite{Finelli:2008jv,De:2014qza,Book-Durrer-2020}, magnetized plasma~\cite{Giovannini-PhysRevD.80.123013,Giovannini:2010ar}, and parity-violating physics~\cite{Alexander:2017bxe,Alexander:2018iwy,Alexander:2019sqb}. Detecting a primordial $V$-mode signal would provide a powerful probe of beyond standard model scenarios, particularly those involving parity-violation in the early Universe~\cite{Lembo:2020ufn,Caloni:2022kwp}. Recent CMB experiments, including the Milano Polarimeter (MIPOL) \cite{Mainini:2013mja}, SPIDER \cite{SPIDER:2017kwu}, and the Cosmology Large Angular Scale Surveyor (CLASS) \cite{Padilla:2019dhz-CLASS,Eimer_2024-CLASS} have placed the first constraints on the angular power spectrum of primordial V-modes, $C_l^{VV}$, marking an important step towards constraining new physics through CMB polarization.

Axions and axion-like particles (ALPs) represent a compelling extension of the standard model of particle physics. Axions were originally proposed
to solve the long-standing strong CP problem in quantum chromodynamics (QCD)~\cite{Peccei:1977hh,Weinberg:1977ma}, 
where their mass $m_a$ and photon coupling $g_{a\gamma\gamma}$ are related~\cite{Marsh:2015xka,OHare:2024nmr}, effectively leaving a single free parameter. However, ALPs encompass a broader class of pseudoscalar particles that naturally arise in string theory~\cite{Sikivie:2006ni,Kawasaki:2013ae,Marsh:2015xka,Sikivie:2020zpn,OHare:2024nmr}. They are also well-motivated dark matter candidates, with their mass $m_{\phi}$ and ALP-photon coupling $g_{\phi\gamma}$ being independent free parameters~\cite{Preskill:1982cy,Abbott:1982af,Dine:1982ah}. 
Several methods have been proposed to constrain the $(m_{\phi},g_{\phi\gamma})$ parameter space of ALPs ~\cite{Sikivie:1983ip,1988-Raffelt.Stodolsky-PRD,HARARI199267,Sikivie:2006ni,Payez:2014xsa,Marsh:2015xka,CAST:2017uph,Fujita:2018zaj,Obata:2018vvr,Fedderke:2019ajk,Zhang:2024dmi,OHare:2024nmr,Taruya:2025bhe}. One of the widely studied mechanism involves axion–photon conversion in the presence of background magnetic fields~\cite{Sikivie:1983ip,1988-Raffelt.Stodolsky-PRD,2001-Deffayet.etal.Zaldarriaga-PRD,Vogel:2017fmc,Sikivie:2020zpn,Mukherjee:2018oeb,Hook:2023smg,Natwariya:2025bnz,Setabuddin:2025vlc}. The conversion of CMB photons into ALPs can induce spectral distortions in the background CMB radiation~\cite{Mukherjee:2018oeb,Hook:2023smg,Cyr:2024sbd}. These studies primarily constrain the resulting deviations in the total intensity from the blackbody spectrum, $\Delta I/I$ using the observational limits from  COBE/FIRAS satellite, as well as forecasts for upcoming or proposed missions with enhanced sensitivity such as PIXIE, SPECTER, BISOU etc~\cite{Chluba:2025wxp} or constraints from linear polarization of CMB~\cite{Fedderke:2019ajk}.

In this {\it work}, we investigate the resonant conversion of ALPs into photons in the presence of a primordial helical magnetic field, which naturally produces circularly polarized (chiral) photons (see Refs.~\cite{1962-Gertsenshtein-JETP,1974-Zeldovich-SJETP,Chen:1994ch,Bastianelli:2004zp,Bastianelli:2007jv,2013-Chen.Suyama-PRD,2012-Dolgov.Ejlli-JCAP,2013-Dolgov.Ejlli-PRD,2020-Fujita.Kamada.Nakai-PRD,2021-Domcke.Garcia-Cely-PRL,Domcke:2022rgu,2022-Kushwaha.etal-MNRAS,2023-Kushwaha.Sunil.Shanki-IJMPD,2023-Palessandro.Rothman-PDU,2023-Dolgov.etal-Universe,Anninos:2024esx,2024-He.Sharma.etal-JCAP,Kushwaha:2025mia,Pappas:2025zld,Domcke:2025qlw} for a similar mechanism of graviton-photon conversion). We focus on the epoch after $\mu$-type spectral distortions ($5\times10^4 \lesssim z \lesssim 2\times 10^6$)~\cite{Cyr:2024sbd,Chluba:2025wxp}, where the excess of such chiral photons generates an additional circularly polarized intensity, which can be probed via the CMB V-modes.
The CMB spectral intensity distortions can manifest either as a reduction in intensity or as an excess of radiation. We focus on the latter, wherein the conversion of ALPs to photons produces excess radiation. 
To the best of our knowledge, this work presents the first investigation of the generation of circularly polarized radiation through the axion–photon conversion in the early Universe, and the first to utilize CMB $V$-mode polarization as a probe to constrain the ALP parameter space (see Refs.~\cite{Yao:2022col,Shakeri:2022usk} for related late-time phenomena)

\section{Axion--photon conversion} 
For high-frequency electromagnetic (EM) waves (photons), where the Hubble parameter $H$ characterizing the expansion rate of the Universe, is much smaller than the frequency of the waves, $H \ll \omega$, the effect of cosmic expansion can be safely neglected~\cite{2021-Domcke.Garcia-Cely-PRL,Kushwaha:2025mia}. 
Therefore, the Lagrangian density describing the axion-photon system in Minkowski spacetime is given by
\begin{equation}\label{lagangian}
\mathcal{L} = - \frac{1}{2} \partial_{\mu}\phi \partial^{\mu}\phi - \frac{1}{2} m_{\phi}^2 \phi^2  - \frac{1}{4} F_{\mu\nu} F^{\mu\nu}  -  \frac{1}{4}g_{\phi\gamma} \phi F_{\mu\nu} \tilde{F}^{\mu\nu} 
\end{equation}
where $\phi$ is the pseudoscalar field (or ALPs) of mass $m_{\phi}$ and $g_{\phi\gamma}$ (with inverse mass dimension) is the coupling function describing the interaction of ALPs with photons. The EM field tensor $F_{\mu\nu}$ is defined by $F_{\mu\nu} = \partial_{\mu} A_{\nu}-\partial_{\nu} A_{\mu}$, with $A^{\mu}$ being the EM vector field, and its dual is defined as $\tilde{F}^{\mu\nu} = \frac{1}{2}\epsilon^{\mu\nu\alpha\beta}F_{\alpha\beta}$. 
The dynamics of ALPs and the EM field 
is described by the Klein-Gordon and Maxwell equations as~\cite{1988-Raffelt.Stodolsky-PRD,2001-Deffayet.etal.Zaldarriaga-PRD,Mirizzi:2007hr,Addazi:2024mii} 
\begin{equation}\label{kg-maxwell-eqns}
    \big(\Box  - m_{\phi}^2\big) \phi = g_{\phi\gamma} \dot{A}_i \, \bar{B}^i~, \; \big(\Box - \omega_{\rm pl}^2 \big)A^i = - g_{\phi\gamma} \dot{\phi} \,  \bar{B}^i
\end{equation}
where $\Box \equiv -\frac{\partial^2}{\partial t^2} + \partial_i \partial_i$ is the d'Alembertian operator, $\dot{}=\partial_t$ and $\bar{B}^i = (\bar{B}_x,\bar{B}_y,\bar{B}_z)$ is the background magnetic field, and $\omega_{\rm pl}=e^2 n_e/m_e$ represents the effective mass of the photon due to plasma effects~\cite{Kushwaha:2025mia}.
The propagating EM waves' gauge field is denoted by $A^i$ and we used $\bar{B}_i = \eta_{ijl} \partial_j \bar{A}_l$, where $\eta_{ijk}$ is the 3D totally antisymmetric symbol.

For EM waves propagating along the $z-$direction with a specific frequency of the modes, 
$A_{\sigma} (z,t) = A_{\sigma} (z) e^{i\omega t}$, the axion-photon system~\eqref{kg-maxwell-eqns} is described by linearized differential equations, which can be written in the matrix form as~\cite{,Mirizzi:2007hr,Addazi:2024mii}
\begin{align}\label{oscillation-eq-matrix}
	\big[(\omega - i\partial_z) \mathbb{I} + \mathcal{M} \big]   \hat{\Psi} (z)  \simeq 0 ~,
\end{align}
where $\mathbb{I}$ is the identity matrix, $ \hat{\Psi} (z) $ is the column vector field, and $\mathcal{M}$ is the mixing matrix given by (see the supplementary material (see appendix \ref{con-prob-derivation} for more details)
\begin{align}\label{mixing-matrix-def}
\hat{\Psi} (z) = \begin{bmatrix}
	A_x (z) \\ A_y (z)\\ \phi (z)
\end{bmatrix}, \quad
  \mathcal{M} = \begin{bmatrix}
        \Delta_{\rm pl} & 0 & \Delta_{\phi\gamma}^x \\
        0 & \Delta_{\rm pl} &  \Delta_{\phi\gamma}^y\\
         \Delta_{\phi\gamma}^x &  \Delta_{\phi\gamma}^y &  \Delta_{\phi}
    \end{bmatrix}~.
\end{align}
The choice $\textbf{A}(t,z) = i (A_x,A_y,0) e^{i\omega t}$ gives the electric field $\textbf{E} = -\partial_t\textbf{A} = \omega (A_x,A_y,0) e^{i\omega t}$, hence real mixing matrix $\mathcal{M}$~\cite{2001-Deffayet.etal.Zaldarriaga-PRD}.
The resonant conversion of ALPs to photons requires the magnetic field to be transverse to the direction of propagation of EM waves, which implies that the magnetic field should be in the $(x-y)$ Cartesian plane, and the conversion is described as the solution of Eq.~\eqref{oscillation-eq-matrix}. Often the magnetic fields are taken to be along $x$ or $y$ direction, reducing the mixing matrix $\mathcal{M}$ to a $2\times2$ matrix (two-level system), describing the conversion of ALPs to photons with polarization parallel to the magnetic field direction, $A_{\parallel}$. This is obvious from the term $\partial_tA_i \bar{B}^i$ in Eq.~\eqref{kg-maxwell-eqns}. Therefore, in this scenario, only linearly polarized EM waves are produced. Furthermore, even if the background magnetic field is in the $(x-y)$ plane and non-helical in nature, the produced EM radiation would always be linearly polarized, which has been extensively investigated in the literature~\cite{Sikivie:1983ip,1988-Raffelt.Stodolsky-PRD,2001-Deffayet.etal.Zaldarriaga-PRD,Mirizzi:2005ng,Mirizzi:2007hr,Mirizzi:2009iz,Mirizzi:2009nq,Addazi:2024mii,Setabuddin:2025vlc}. In this context, an interesting question arises: \textit {how are the polarization properties of photons affected by the presence of a helical background magnetic field?} In particular, can such a field lead to the production of chiral photons, rather than only linearly polarized ones?
Chiral photons are of great importance in light of CMB V-mode measurements and would provide an independent insight in probing the parity violation in the early Universe~\cite{Kushwaha:2025mia}. This is because the CMB V-mode experiments measure the intensity difference between both the helicity modes of the chiral photons, which probe a key quantity, i.e., net circular polarization. 
Helical magnetic fields are particularly interesting to probe the parity violation in the Universe due to their different geometrical properties from the non-helical ones, as they describe the twist or linkage of the magnetic field lines~\cite{2011-Durrer.Hollenstein.Jain-JCAP,2012-Byrnes.etal-JCAP,2013-Durrer.Neronov-Arxiv,2016-Subramanian-Arxiv,2018-Sharma.Subramanian.Seshadri.PRD,2020-Kushwaha.Shankaranarayanan-PRD,2021-Kushwaha.Shankaranarayanan-PRD}, and also possess a non-vanishing helicity density (see appendix \ref{appsec-magnetic-field} for more details). 

To investigate the effects of axion-photon conversion on CMB, we consider the resonant conversion of ALPs to photons to occur before the decoupling epoch, specifically during redshift $1100 < z < 10^4$, and leading to distortions in the CMB thermal blackbody spectrum. During the radiation-domination (RD) epoch i.e., $z>z_{\rm dec}$, the Hubble parameter redshifts as $H\sim (1+z)^2$, and after the decoupling epoch (i.e., $z_{\rm dec} \simeq 1100$) till today, i.e., during the epoch of matter-domination, $H (z) \sim (1+z)^{3/2}$. At decoupling, $H_{\rm dec} = H_0\left( T_{\rm dec}/T_0 \right)^{3/2}$, where $T_{\rm dec} = 0.26 \, {\rm eV}$ is the temperature of the Universe at decoupling and $H_0 = 2.13\times 10^{-33} \, {\rm eV}$.
During this epoch, the refractive index of the medium for high-frequency photons is determined by the electron number density, as well as the presence of neutral hydrogen and helium. Furthermore, the electron number density during this epoch is given as $n_e (z)  = n_{b0} (1+z)^3 \, X_e (z)$, where $n_{b0} = 0.251 \, {m}^{-3} $ 
is the baryon number density today and $X_e(z)$ is the ionization fraction. This gives the plasma frequency as $\omega_{\rm pl} (z) \simeq 27.67 \, {\rm Hz} \, (1+z)^{3/2} X_e^{1/2} (z)$. The resonant conversion occurs at a redshift $z_{\rm res}$ when $m_{\phi}$ becomes approximately equal to plasma frequency $\omega_{\rm pl} (z_{\rm res})$~(see appendix \ref{appsec-mass} for more details). 
Therefore, during the epoch of our interest, i.e., $z_{\rm res}\in [1100,10^4]$, the mass range of ALPs can be obtained as $m_{\phi} \in [2.58\times10^{-10}, 1.82\times 10^{-8}] \, {\rm eV}$.
\begin{figure}[t!]
%
\includegraphics[height=2.3in,width=3in]{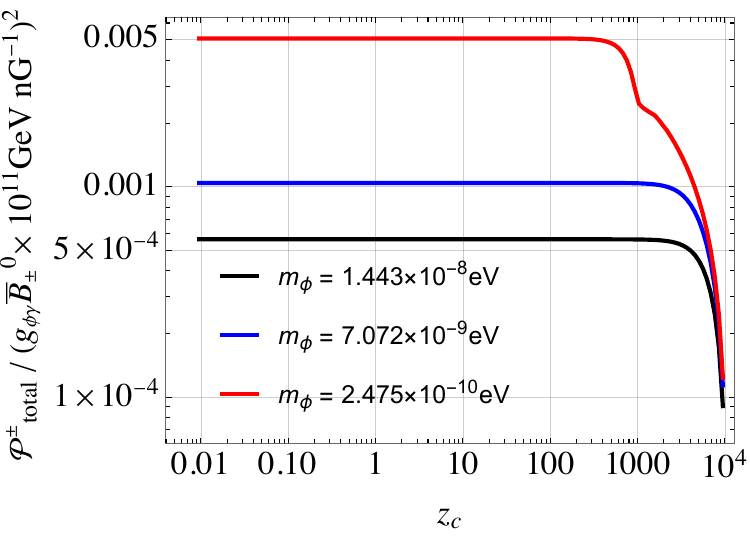} 
%
%
\caption{The conversion probability for three typical ALPs masses going through resonance between $1100 \lesssim z_{\rm res} \lesssim  10^4$ for CLASS observation frequency, $\omega_0 = 2\pi\times40 \, {\rm GHz}$.}
\label{fig-con-prob}
\end{figure}
The conversion probability of ALPs to photons in the redshift interval $[z_j,z_{j+1}]$ is given by (see appendix \ref{con-prob-derivation} for more details)
\begin{align}\label{con-prob-pm-z-fin-num}
P^{\pm}_{\phi\rightarrow\gamma^{\pm}} (z_j,z_{j+1}) = \frac{\pi}{8} \left( \frac{g_{\phi\gamma}}{{\rm GeV^{-1}}} \right)^2  \left( \frac{ | \bar{B}_{\pm}^0 | }{{\rm nG}} \right)^2 \left( \frac{\rm eV}{4.96\times 10^{19}} \right)^2 \nonumber\\ \times \frac{(1+z_c)^3}{H_c\,|\Delta_{\rm osc}' (z_c) | } \Big|\,{\rm Erf}[(1+i)\alpha_j] - {\rm Erf}[(1+i)\alpha_{j+1}] \,\Big|^2 \, ,
\end{align}
where ${\rm Erf}$ is the error function and the dimensionless parameters $\alpha_{s=j,j+1}$ are given as
\begin{subequations}\label{alpha}
    \begin{align}
    \alpha_s &= \frac{\Delta_{\rm osc} (z_c) + (z_s -z_c) \, |\Delta_{\rm osc}'(z_c)|}{2 \sqrt{H_c (1+z_c) \, |\Delta_{\rm osc}'(z_c)|}}  ~, \\  \alpha_{j+1} &= \frac{\Delta_{\rm osc} (z_c) + (z_{j+1} -z_c) \, |\Delta_{\rm osc}'(z_c) | }{2 \sqrt{H_c (1+z_c) \, | \Delta_{\rm osc}'(z_c) | }} ~.
\end{align}
\end{subequations}
where subscript `c' refers to the quantities evaluated at the mid-point of each redshift interval, $z_c=(z_j+z_{j+1})/2$ (see appendix \ref{con-prob-derivation} for plots of different terms in Eq.~\eqref{con-prob-pm-z-fin-num}). The discretization scheme is provided by the decreasing sequence of redshift, $z_{j+1} = z_j (1-\epsilon)$~\cite{Addazi:2024mii}. We take $\epsilon=0.1$, which sets the physical condition that the ALP to photon mixing propagation length remains smaller than the comoving Hubble radius $l \simeq \epsilon\,\mathcal{H}^{-1}$, and the comoving distance during each redshift interval can be approximately given by $\Delta l \simeq \epsilon z_c/H_c$~\cite{Addazi:2024mii}.
In Eq.~\eqref{con-prob-pm-z-fin-num}, $\bar{B}^0_{\pm}$ refers to the present day primordial magnetic field strength~(see Refs.~\cite{1994-Kronberg-Rept.Prog.Phys.,1996-Beck.etal-ARAA,2001-Grasso.etal-PhyRep,2002-Widrow-Rev.Mod.Phys.,2004-Vallee-NAR,2004-Gaensler.Beck.etal-AstroRev,2004-Giovannini-IJMPD,2007-Barrow.etal-PhyRep,2011-Kandus.Kunze.Tsagas-PhysRept,2013-Durrer.Neronov-Arxiv,2016-Subramanian-Arxiv}) of each helicity component (`$+$' for right-handed and `$-$' for left-handed), which is constrained by CMB Planck, for maximally helical case $\bar{B}^0 \sim {\rm nG}$~\cite{Planck:2015zrl,2016-Subramanian-Arxiv}. 
In the literature, often the combination of parameters $g_{\phi\gamma}\bar{B}^0$ is constrained from photon-axion conversion for a range of $m_{\phi}$~\cite{Mirizzi:2009nq}, which renders these constraints free from uncertainties in the value of the primordial magnetic field strength. 
Thus, using the CMB constraints on cosmological magnetic field~\cite{2013-Durrer.Neronov-Arxiv,Planck:2015zrl,2016-Subramanian-Arxiv} allows us to constrain the ALP parameter space $(g_{\phi\gamma},m_\phi )$.

Before closing this section, let us obtain the total conversion probability by interpolating Eq.~\eqref{con-prob-pm-z-fin-num} up to any redshift $z$ as $\mathcal{P}^{\pm}_{\rm total} (z_i) = \sum_{j=1}^{j=i} P^{\pm}_{\phi \rightarrow \gamma^{\pm}} (z_j)$. \ref{fig-con-prob} shows the total conversion probability for three typical masses, $m_{\phi}\simeq 1.443\times 10^{-8} \, {\rm eV} \, , 7.072\times 10^{-9} \, {\rm eV} \, , 2.475\times 10^{-10} \, {\rm eV}$ going through resonance at redshifts $z_{\rm res}\simeq 8550 \, ,5314.41 \, , 1094.18$, respectively. From \ref{fig-con-prob}, we find that the total conversion probability peaks at the corresponding $z_{\rm res}$ and becomes constant till the present epoch, implying no further resonant conversion.

\section{Excess chiral radiation}
The CMB photons were in thermal equilibrium with the primordial plasma before the epoch of decoupling and are well described today by a Bose-Einstein distribution function $f_{\gamma}^{\rm eq} = 1/(e^{\omega/T}-1)$. The intensity of radiation is related to the distribution function as, $ I (\omega,T) = \frac{\omega^3}{4\pi^2} f_{\gamma} (\omega,T)$, where we omit an overall factor of $2$ corresponding to two modes of polarization of photon~\cite{Book-Baumann-2022}, as we are treating them separately. Due to axion-photon conversion, some ALPs will be converted to photons, which would affect the spectrum of CMB photons, causing the distortions of the thermal blackbody spectrum~\cite{Chluba:2025wxp}. We can quantify this distortion by estimating the difference $\Delta f_{\gamma} = f_{\gamma} - f^{\rm eq}_{\gamma}$, 
where $f_{\gamma}$ is the total photon distribution. Assuming, before the onset of the resonant conversion of ALPs to photons (during $1100\lesssim z_{\rm res} \lesssim 10^4$), the distortion in the spectrum is zero, i.e., $\Delta f_{\gamma} = 0$ or the corresponding intensity distortion $\Delta I = 0$, which essentially means that the distortions are only caused by this mechanism. 
The distribution function of photons allows us to define the dimensionless energy density parameter as
$\Omega_{\gamma} (\omega,T) = \frac{1}{\rho_c} \frac{\omega^4}{\pi^2} f_{\gamma} (\omega,T)$, where $\rho_c$ is the critical energy density of the Universe. 
Upon integrating over all frequencies, we get $\Omega_{\gamma} (T) = \frac{\pi^2}{15} \frac{T^4}{\rho_c}$.
Now using the relation $\Omega_{\gamma} (T) = \frac{\rho_{\gamma}}{\rho_c}=\frac{\pi^2}{15} \frac{T^4}{\rho_c}$, we can obtain the distortion in the intensity of radiation due to excess chiral photons (of each helicity) as 
\begin{align}
\Delta I^{\pm} (\omega, T) = \frac{\pi^2}{60} \left(\frac{T}{\omega} \right)^4 \omega^3 \frac{\Omega^{\pm}_{\gamma} (\omega,T)}{\Omega_{\gamma} (T)},
\end{align}
where $\Omega^{\pm}_{\gamma} (\omega,T)$ is the energy density per logarithmic frequency interval of the excess chiral photons, which is solely sourced by ALPs to the photon conversion mechanism, and is proportional to the conversion probability. 
At a given redshift and for each helicity mode of photons, the energy density per logarithmic frequency interval of the excess chiral photons $\Omega^{\pm}_{\gamma} (\omega,T)$ is given by~\cite{Addazi:2024mii}
\begin{align}\label{excess-rad}
    \Omega_{\gamma}^{\pm} (\omega,T) = \Omega_{\rm ALP} (\omega,T) \, \mathcal{P}_{\rm total}^{\pm} (\omega,z)
\end{align}
where $\Omega_{\rm ALP} (\omega,T)$ is the energy density per logarithmic frequency interval of ALPs, at temperature $T = T_0 (1+z)$. 
Thus, the fractional change in the intensity of excess radiation with respect to the CMB photons in thermal equilibrium (blackbody spectrum) can be determined by the relation\footnote{In our work~\cite{Kushwaha:2025mia}, we considered distortions relative to the peak intensity, which introduce a frequency-dependent suppression, $\frac{\Delta I^{\pm}}{I^{\rm eq}_0} = \left( \frac{\omega_0 }{\omega_0^{\rm eq}}\right)^3 \frac{\Delta f_{\gamma}^{\pm}}{f^{\rm eq}_{\gamma,0}}$, where $\omega_0^{\rm eq} \simeq 2.82 \, T_{0}$. As a result, the intensity distortion at lower frequencies (e.g., $\nu_0 = 33\, \& \, 40 \, {\rm GHz}$) is suppressed and corresponds to reduced conversion probabilities. Correcting for this effect, as done in the present work, yields more stringent constraints at low frequencies than those reported in Ref.~\cite{Kushwaha:2025mia}.}, $\frac{\Delta I^{\pm}}{I^{\rm eq}_0} =\frac{\Delta f_{\gamma}^{\pm}}{f^{\rm eq}_{\gamma,0}}$, where $I^{\rm eq}_0$ is the intensity of CMB photons corresponding to $f_{\gamma,0}^{\rm eq}$ at the present epoch, i.e., $T_{0} \simeq 2.726 \, K \simeq 354 \, {\rm GHz}$ 
~\cite{2024-Kushwaha.Jain-PRD}.
Thus, the ratio at the present epoch is given by 
\begin{align}\label{excess-intensity-1}
    \frac{\Delta I^{0,\pm}}{I_0^{\rm eq}} = \frac{\pi^4}{15} \left(\frac{T_0}{\omega_0}\right)^4  \frac{1}{f^{\rm eq}_{\gamma,0} }\frac{\Omega_{\rm ALP} (\omega_0,T_0)}{\Omega_{\gamma} (T_0)} \, \mathcal{P}_{\rm total}^{\pm} (\omega,z)
\end{align}
where we used the scaling relations $\Omega_{\rm ALP}, \Omega_{\gamma} \sim (1+z)^4$. For simplicity of the calculations and to obtain the conservative estimates, we assume frequency-independent ALP abundance, such that the energy density of ALPs $\Omega_{\rm ALP} (T,\omega) = \int d\ln{\omega} \, \Omega_{\rm ALP} (T,\omega) \simeq \Omega_{\rm ALP} (T) $, and evolves as $\Omega_{\rm ALP}, \Omega_{\gamma} \sim (1+z)^4$. We further define the frequency-independent parameter fixed at present epoch as, $ \alpha= \frac{\Omega_{\rm ALP} (\omega_0,T_0)}{\Omega_{\gamma} (T_0)}$, where $\alpha$ is constrained as $\alpha = 0.23\, \Delta N_{\rm eff} \lesssim 0.06$~\cite{Planck:2015zrl,Addazi:2024mii}.
Thus, the total excess of chiral photons due to the axion-photon conversion mechanism determined by Eq.~\eqref{excess-rad} would contribute to the excess of background EM radiation which can be quantified in terms of the excess intensity $\Delta I  (\omega,T)$ of photons in the frequency range $\omega$ to $\omega + d\omega$.
This provides the relation to estimate the total excess intensity due to both polarizations at the present epoch, $\Delta I = \Delta I_0^+ + \Delta I_0^-$, at a given frequency $\omega_0$ as 
\begin{align}\label{excess-intensity-final}
    \frac{\Delta I}{I_0^{\rm eq}} &= \frac{\pi^4}{15} \left(\frac{T_0}{\omega_0}\right)^4\frac{1}{f^{\rm eq}_{\gamma,0} }\frac{\Omega_{\rm ALP} (\omega_0,T_0)}{\Omega_{\gamma} (T_0)} \mathcal{P}_{\rm total} (\omega,z)~,
\end{align}
where the total conversion probability, $\mathcal{P}_{\rm total} (\omega,z_i) = \sum_{j=1}^{j=m} (P^+_{\phi \rightarrow \gamma^{\pm}} (z_j)+P^-_{\phi \rightarrow \gamma^{\pm}} (z_j))$, 
in the above equation can be obtained by
\begin{align*}\label{tot-con-prob}
    \mathcal{P}_{\rm total} (\omega,z_m) 
    = \frac{\pi}{8 \, } \left( \frac{g_{\phi\gamma}}{{\rm GeV^{-1}}} \right)^2  \left( \frac{ \mathcal{B}_0 }{\sqrt{2}\,{\rm nG}} \right)^2 \left( \frac{\rm eV}{4.96\times 10^{19}} \right)^2 \nonumber\\ 
    \times \sum_{j=1}^{j=m}\frac{(1+z_c)^3}{H_c\,|\Delta_{\rm osc}' (z_c) | } \left| \, {\rm Erf}[(1+i)\alpha_j] 
    - {\rm Erf}[(1+i)\alpha_{j+1}] \,\right|^2 
\end{align*}
%
%
where $\mathcal{B}_0 
= \sqrt{ 2 (|B_{0,+}|^2 + |B_{0,-} |^2 )}$ (see appendix \ref{appsec-magnetic-field} for more details).

\section{Constraints from the V-mode measurements of CMB circular polarization}
The properties of the EM radiation are described by the Stokes parameters, which are usually defined in terms of two orthogonal components of the electric field $E$. The total intensity of the radiation (photons) is determined by $I$-Stokes parameter as $I = I_0 + \Delta I_{\rm CMB} + \Delta I_{\phi \rightarrow \gamma}$, where $I_0$ is the peak intensity of the background CMB photons, $\delta I_{\rm CMB}$ is the intensity corresponding to the distortion in the CMB spectrum due to various other mechanisms~\cite{Kosowsky:1994cy,Hu:1997hv,Komatsu:2022nvu,Chluba:2025wxp}, and $\Delta I_{\phi \rightarrow \gamma}$ is the intensity of the excess photons due to axion-photon conversion\footnote{In axion--photon conversion, the intensity distortion can arise in two ways: (i) a \emph{reduction} due to CMB photons converting into axions, and (ii) an \emph{excess} due to axions converting into photons. In this work, we focus on the latter and assume that axions are more abundant than photons during the epoch of interest, without specifying the underlying beyond-standard-model (BSM) mechanism responsible for this abundance. Consequently, constraining axion parameters from the induced $V$-mode can equivalently be interpreted as constraining this assumption about the initial axion abundance.}. Similarly, we can split the $V-$Stokes parameter as $V = V_0 + \Delta V$, where at the background level $V_0 = 0$ as the circular polarization is not produced via Thomson scattering, and $\Delta V$ is the contribution due to the axion-photon conversion mechanism. 
Assuming that the axion-photon conversion is the only source of excess circularly polarized radiation, in flat Friedmann–Lemaître–Robertson–Walker spacetime, we find the Stokes parameters~\cite{Kosowsky:1994cy,Kamionkowski:1996ks,Alexander:2018iwy,Alexander:2019sqb,Kushwaha:2025mia} as $ \Delta I = \Delta I_{\phi\rightarrow \gamma}  = \frac{| \Delta E_+|^2}{a^2} \left( 1 + \delta_A \right)$ and $\Delta V  = \frac{| \Delta E_+|^2}{a^2} \left( 1 - \delta_A \right)$, where $\delta_A = (|A_-|/|A_+|)^2$~\cite{2024-Kushwaha.Jain-PRD,Kushwaha:2025mia} and $a(t)$ is the scale factor. 
Using the definition of the chirality parameter of the photons $\Delta \chi_\gamma = \frac{1-\delta_A}{1+\delta_A}$, gives an important relation $\Delta V = \Delta \chi_{\gamma} \cdot \Delta I$.
%
Now, using Eqs.~\eqref{con-prob-pm-z-fin-num} and \eqref{excess-rad} along with the definition of the chirality parameter of the background helical magnetic field, $\frac{\langle |\bar{B}^0_{+}|^2 \rangle - \langle |\bar{B}^0_{-} |^2 \rangle}{\langle |\bar{B}^0_{+}|^2 \rangle + \langle |\bar{B}^0_{-} |^2 \rangle} = \frac{1-\delta_B}{1+\delta_B} = \Delta \chi_B$ (see the appendix~\ref{appsec-magnetic-field} for more details), we can straightforwardly show that $\Delta \chi_{\gamma} = \Delta \chi_B$. %
This is an important result which indicates that the net chirality of the produced photons is solely determined by the chirality of the background magnetic field $\Delta\chi_B$. 

Stokes-$V$ and Stokes-$I$ parameters both have units of intensity. However, because the CMB follows an almost perfect blackbody spectrum, its intensity can be uniquely expressed in terms of temperature. Consequently,
the fluctuations (anisotropies) observed in the CMB experiments are measured in units of temperature~\cite{Alexander:2017bxe,2002-Staggs.etal-proceeding,Kushwaha:2025mia}. We define the $V$-mode in temperature units through the relation $\frac{V}{ I} = \frac{ V_T}{ T}$, and since CMB experiments measure the $V$-mode in temperature units, $V_T$ serves as our observable quantity. The power spectrum of the $V$-mode can be obtained by the two-point correlation function as~\cite{Alexander:2017bxe}
\begin{align}
	\Bigg\langle \frac{\Delta V_T}{T_{0}} \, \frac{\Delta V_T}{T_{0}} \Bigg\rangle  = \Bigg\langle \left(\Delta \chi_{B} \right)^2 \, \frac{\Delta I}{I^{\rm eq}_0} \, \frac{\Delta I}{I^{\rm eq}_0}  \Bigg\rangle ~~.
\end{align}
This relation is central to our analysis: the left-hand side (LHS) is fixed by CMB $V$-mode measurements, while the right-hand side (RHS) depends on the ALP parameters that we aim to constrain. From experiments\footnote{To obtain conservative constraints, we assume a global mapping between the total excess circularly polarized intensity generated by axion-photon conversion and the observed $V$-mode power spectrum. Accordingly, we use the measurement at a specific multipole chosen where the power spectrum amplitude is minimal to derive upper limits. A more rigorous treatment would require evolving the generated circular polarization through the full Boltzmann hierarchy, including transfer functions, line-of-sight integrals, and visibility-weighted projections, analogous to the standard analysis of CMB temperature and polarization anisotropies (see, e.g., Refs.~\cite{Kosowsky:1994cy,Hu:1997hv,Kamionkowski:2015yta,Book-Durrer-2020,Komatsu:2022nvu}).}, the measured quantity is
$\langle \Delta V_T \, \Delta V_T \rangle = \frac{l (l+1)}{2\pi} C_l^{VV} T_{0}^2$ and we use the most recent bounds on the $V$-mode angular power spectrum to derive upper limits on the ALP parameter space $(g_{\phi\gamma}, m_{\phi})$.
\begin{figure}[t!]
\includegraphics[height=2.3in,width=3in]{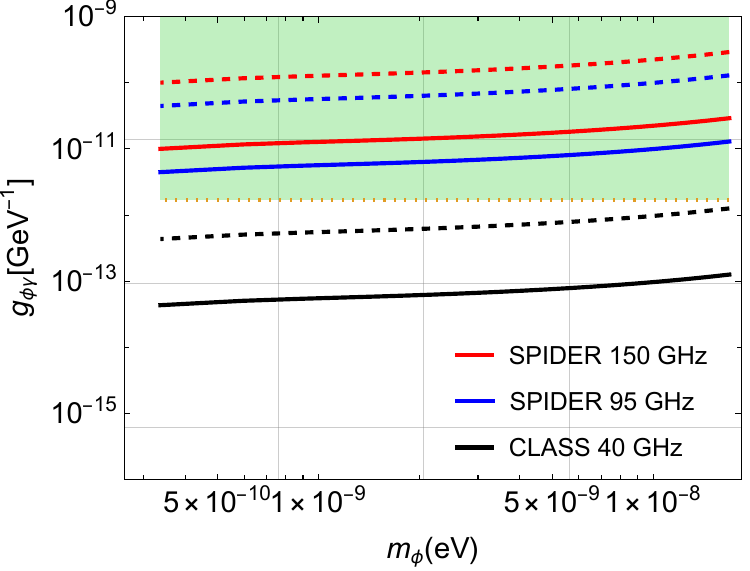} 
\caption{The resulting constraints plot on the ALPs parameter space $(g_{\phi\gamma},m_\phi)$ from various CMB $V$-mode measurements. For simplicity, we have fixed $\Delta \chi_B = 1$, and the solid and dashed curves represent $\mathcal{B}_{0}=1\, {\rm nG}$ and $\mathcal{B}_{0}= 0.1\, {\rm nG}$, respectively. The green shaded region shows the current constraints from magnetic white dwarf (MWD) polarization \cite{Benabou:2025jcv}. 
}
\label{fig-constraints-plot}
\end{figure}
Recently, there has been an improvement in the $V$-mode measurements. For example, the strongest current constraint comes from the ground-based CLASS experiment, $ \frac{l (l+1)}{2\pi} C_l^{VV} T_{0}^2 < 0.023 \, {\mu K^2}$~\cite{Eimer_2024-CLASS} at frequency $\nu_0 = 40 \,{\rm GHz}$ (where $\nu_0 = \omega_0/2\pi$) over the multipole range $5 < l < 125$~\cite{Eimer_2024-CLASS}.
Previously, the balloon-borne experiment SPIDER~\cite{SPIDER:2017kwu} constraints were at higher multipole (in the range $33 < l < 307$) at frequencies $95\, {\rm GHz}$ and $150\,{\rm GHz}$ are $\leq 783 \, {\mu K^2}$ and $\leq 141 \, {\mu K^2}$, respectively. Also, see Ref.\cite{Raffuzzi:2024wyh} for projected sensitivity from LiteBIRD-like and Simons Observatory (SO).

We present our derived constraints on the ALP parameter space $(g_{\phi\gamma}, m_{\phi})$ in \ref{fig-constraints-plot} using the latest $V$-mode measurements from SPIDER and CLASS, alongside existing bounds from magnetic white dwarf polarization (green shaded region
)~\cite{Dessert:2022yqq,Benabou:2025jcv}. Notably, the CLASS results provide the most stringent limits, probing regions of the parameter space that have remained unconstrained so far. Our work demonstrates, for the first time, that CMB $V$-mode polarization measurements can serve as a competitive probe of ALP physics, opening up improved sensitivity to regions of $(g_{\phi\gamma}, m_\phi)$, otherwise inaccessible to previous searches.

\section{Discussion}
 
In this work, we have proposed a novel avenue to constrain the parameter space of ALPs by exploiting the $V$-mode measurements of the circular polarization of CMB. Previous studies have mainly focused on spectral distortions in the total intensity of the CMB blackbody spectrum arising from the axion–photon conversion. In contrast, our analysis demonstrates that in the presence of a helical cosmological magnetic field before the epoch of recombination, resonant axion–photon conversion can generate an excess of circularly polarized radiation. This mechanism provides a direct physical connection between $V$-mode measurements and the underlying ALP parameters, offering a new and complementary observational window to probe ALP–photon interactions beyond the conventional intensity-based methods.

Our results highlight that, under the optimistic scenario with a maximally helical magnetic field of strength $\sim {\rm nG}$, various current and upcoming CMB polarization experiments such as CLASS are sensitive to the unexplored ALP mass range of $10^{-10}$–$10^{-8}\,\mathrm{eV}$ and provide the most stringent constraints to date. Our constraints are particularly relevant, as this mass window remains poorly constrained by existing astrophysical and laboratory searches. 
Consequently, $V$-mode polarization observations represent a powerful and unique probe that can bridge this gap and potentially uncover new physics associated with light ALPs and the helical nature of cosmic magnetic fields.

Given the fact that CMB circular polarization has not received the required attention so far,  current observational bounds on the 
$V$-mode measurements remain relatively weak. However, with targeted efforts and the development of future CMB experiments designed with dedicated sensitivity to CMB circular polarization, the existing constraints could be significantly improved. Such advancements would not only enhance our ability to test fundamental physics,  such as axion–photon interactions and parity-violating processes in the early Universe, but also establish the 
$V$-mode polarization as a powerful and complementary observable in CMB cosmology and could also be equally helpful in constraining beyond standard model physics.

\section*{Acknowledgements}
The work of A.K. was supported by the Japan Society for the Promotion of Science (JSPS) as part of the JSPS Postdoctoral Program (Grant Number: 25KF0107). A.K. thanks Ranjan Laha, Poonam Mehta, and Suvodip Mukherjee for the insightful discussions. We thank Yashi Tiwari and Ujjwal Upadhyay for interesting comments and questions. R.K.J. would like to acknowledge financial support from the Indo-French Centre for the Promotion of Advanced Research (CEFIPRA) for support of the proposal 6704-4 under the Collaborative Scientific Research Programme, IISc Research Awards 2024 and SERB, Department of Science and Technology, GoI through the MATRICS grant~MTR/2022/000821.

\onecolumngrid
\appendix

\section{Derivation of the conversion probability}
\label{con-prob-derivation}
In this appendix, we derive the conversion probability for the axion-photon system, which is described by the matrix equation~\cite{Mirizzi:2007hr,Addazi:2024mii}
\begin{align}\label{appeq-oscillation-eq-matrix}
	\left[ (\omega - i\partial_z) \mathbb{I} + \mathcal{M} \right]   \hat{\Psi} (z)  \simeq 0 ~~,
\end{align}
where $\mathbb{I}$ is the identity matrix, $\mathcal{M}$ is analogous to mixing matrix and $\hat{\Psi} (z)$ is the column vector field, given by
\begin{align}\label{appeq-mixing-matrix-def}
\hat{\Psi} (z) = \begin{bmatrix}
	A_x (z) \\ A_y (z)\\ \phi (z)
\end{bmatrix}, \quad
  \mathcal{M} = \begin{bmatrix}
        \Delta_{\rm pl} & 0 & \Delta_{\phi\gamma}^x \\
        0 & \Delta_{\rm pl} &  \Delta_{\phi\gamma}^y\\
         \Delta_{\phi\gamma}^x &  \Delta_{\phi\gamma}^y &  \Delta_{\phi}
    \end{bmatrix} ~~.
\end{align}
We treat the quantities $\Delta_{\rm pl} = -\frac{\omega^2_{\rm pl}}{2\omega}$ and $\Delta_{\phi} = -\frac{m_{\phi}^2}{2\omega}$ as unperturbed 
and the quantities describing the interaction of ALPs with background magnetic field, $\Delta_{\phi\gamma}^x = \frac{1}{2}g_{\phi\gamma}\bar{B}_x$ and $\Delta_{\phi\gamma}^y =\frac{1}{2}g_{\phi\gamma}\bar{B}_y $ as first order in perturbations. This decomposition into the perturbation is convenient to solve the equations perturbatively, and is well justified because during the epoch of our interest (i.e., CMB), the magnetic field strength is small~\cite{2013-Durrer.Neronov-Arxiv,2016-Subramanian-Arxiv}. Furthermore, we have verified that for the parameter range of $g_{\phi\gamma}$, the product $g_{\phi\gamma}\bar{B}$ is subdominant with respect to the diagonal entries, which will be quantified later in this section.
To solve Eq.~\eqref{appeq-oscillation-eq-matrix}, we follow Ref.~\cite{Mirizzi:2007hr,Addazi:2024mii} and rewrite the matrix equation as
\begin{align}\label{appeq-oscillation-eq-matrix-K}
	\partial_z   \hat{\Psi} (z)  \simeq i \mathcal{K} \hat{\Psi} (z) \qquad \text{where} \qquad  \mathcal{K} = \mathcal{K}_0 + \delta \mathcal{K}\, ,
\end{align}
where $ \mathcal{K}_0$ is the diagonal matrix corresponding to the unperturbed quantities, and $\delta  \mathcal{K}$ is the perturbative matrix which is considered to be a small perturbation due to the interaction of ALPs and the background magnetic field, which are given by
\begin{align}\label{appeq-K-def}
  \mathcal{K}_0 = \begin{bmatrix}
       \omega + \Delta_{\rm pl} & 0 & 0 \\
        0 & \omega + \Delta_{\rm pl} &  0\\
         0 &  0 & \omega + \Delta_{\phi} 
    \end{bmatrix}
   \quad \text{and}
   \quad \delta \mathcal{K} = \begin{bmatrix}
        0 & 0 & \Delta_{\phi\gamma}^x \\
        0 & 0 &  \Delta_{\phi\gamma}^y\\
         \Delta_{\phi\gamma}^x &  \Delta_{\phi\gamma}^y & 0
    \end{bmatrix} ~~.
\end{align}
The solutions of Eq.~\eqref{appeq-oscillation-eq-matrix-K} can be given in terms of the conversion matrix $ \mathcal{U} (\ell,\ell_0) $ by solving the following equation~\cite{Mirizzi:2007hr,Addazi:2024mii}
\begin{align}\label{psi-equation}
  \partial_z \left( e^{-i\int_{\ell_0}^{\ell}d\ell' \, \mathcal{K}_0 (\ell') } \mathcal{U} (\ell,\ell_0) \right)  = i e^{-i\int_{\ell_0}^{\ell} d\ell' \, \mathcal{K}_0 (\ell') } \, \delta \mathcal{K} (\ell)  \mathcal{U} (\ell,\ell_0) 
\end{align}
where the conversion matrix describes the conversion of ALPs to photons after travelling a distance $\ell-\ell_0$. The solution describing the axion-photon oscillation is given as $\hat{\Psi} (\ell) = \mathcal{U} (\ell,\ell_0) \, \hat{\Psi} (\ell_0)$, where the conversion matrix take the following general form up to leading order in perturbation 
\begin{align}\label{U-sol}
   \mathcal{U} (\ell,\ell_0)   = e^{i\int_{\ell_0}^{\ell}d\ell' \, \mathcal{K}_0 (\ell') } + i e^{i\int_{\ell_0}^{\ell}d\ell'' \, \mathcal{K}_0 (\ell'') } \int_{\ell_0}^{\ell} d\ell'  e^{-i\int_{\ell_0}^{\ell'} d\ell'' \, \mathcal{K}_0 (\ell') } \, \delta \mathcal{K} (\ell')  e^{i\int_{\ell_0}^{\ell'} d\ell'' \, \mathcal{K}_0 (\ell') } + \mathcal{O} \left( (\delta \mathcal{K})^2 \right)
\end{align}
As we know, the above conversion matrix describes the axion-photon oscillation, therefore, by choosing the initial conditions at $\ell_0$ such as $A_{x,y}(\ell_0)=0$ (i.e., only axion at the beginning) and $\phi (\ell_0)=0$ (i.e., only photons at the beginning) can be applied to study the axion-to-photon and photon-to-axion conversions, respectively. Since we are interested in axion-photon conversion, the solutions are given as
\begin{subequations}\label{U-sol-xy}
    \begin{align}
         A_x (\ell) &= \mathcal{U}_{13} (\ell,\ell_0) \, \phi (\ell_0)  =  \left( \frac{i}{2} g_{\phi\gamma} \int_{\ell_0}^{\ell} d\ell' \bar{B}_x (\ell') \, e^{-i\int_{\ell_0}^{\ell'} d\ell'' \, \left[ \Delta_{\rm pl}(\ell'') - \Delta_{\phi}(\ell'') \right] } \right) \, \phi (\ell_0) \\
   A_y (\ell) &= \mathcal{U}_{23} (\ell,\ell_0) \, \phi (\ell_0)  =  \left( \frac{i}{2} g_{\phi\gamma} \int_{\ell_0}^{\ell} d\ell' \bar{B}_y (\ell') \, e^{-i\int_{\ell_0}^{\ell'} d\ell'' \, \left[ \Delta_{\rm pl}(\ell'') - \Delta_{\phi}(\ell'') \right] } \right) \, \phi (\ell_0) ~.
    \end{align}
\end{subequations}
Combining these equations allows us to give the corresponding helical components as
\begin{align}\label{Apm-sol-def}
   A_{\pm} (\ell) =  \frac{A_x (\ell) \mp i A_x (\ell) }{\sqrt{2}} \, \phi (\ell_0) &= \frac{ \mathcal{U}_{13} (\ell,\ell_0) \mp i \, \mathcal{U}_{23} (\ell,\ell_0) }{\sqrt{2}} \, \phi (\ell_0)
\end{align}
which further allows us to define the conversion matrix for the ALPs to chiral photons conversion as
\begin{align}\label{Apm-Upm-def}
   A_{\pm} (\ell) =  \mathcal{U}^{\pm} (\ell,\ell_0)  \, \phi (\ell_0), \quad \text{where} \quad   \mathcal{U}^{\pm} (\ell,\ell_0) = \frac{ \mathcal{U}_{13} (\ell,\ell_0) \mp i \, \mathcal{U}_{23} (\ell,\ell_0) }{\sqrt{2}} 
\end{align}
and we have used the helicity relation $  A_{\pm} = \frac{A_x \mp i A_y}{\sqrt{2}} $ and $ \bar{B}_{\pm} = \frac{\bar{B}_x \mp i \bar{B}_y}{\sqrt{2}}$\cite{Kushwaha:2025mia}.
Thus, the solution for the conversion of ALPs to chiral photons is given by 
\begin{align}\label{Apm-sol}
  A_{\pm} (\ell)  =  \left( \frac{i}{2} g_{\phi\gamma} \int_{\ell_0}^{\ell} d\ell' \bar{B}_{\pm} (\ell') \, e^{-i\int_{\ell_0}^{\ell'} d\ell'' \, \left[ \Delta_{\rm pl}(\ell'') - \Delta_{\phi}(\ell'') \right] } \right) \, \phi (\ell_0) ~~.
\end{align}
The conversion probability for the individual helicity mode of chiral photons can be calculated as
\begin{align}\label{ap-con-prob-pm-generic}
P^{\pm}_{\phi\rightarrow\gamma^{\pm}} (\ell)=  \frac{|A_{\pm} (\ell)|^2}{|\phi (\ell_0)|^2}  =   \frac{1}{4} g^2_{\phi\gamma} \int_{\ell_0}^{\ell} d\ell_1 \int_{\ell_0}^{\ell} d\ell_2 \left(\bar{B}_{\pm} (\ell_1) \, \bar{B}_{\pm}^* (\ell_2) \right) \, e^{-i\int_{\ell_2}^{\ell_1} d\ell'' \, \left[ \Delta_{\rm pl}(\ell'') - \Delta_{\phi}(\ell'') \right] } ~~.
\end{align}
Note that the above solution is generic in the sense that, it holds for the case where there are inhomogeneities in the magnetic field and plasma at a given redshift $z$ (from hereafter, we refer to $z$ as redshift). The coordinates/lengths $\ell$ used are physical quantities and can be related to comoving coordinates $l$ as $\ell = l/(1+z)$. The conversion probability in terms of comoving length scales is given by 
\begin{align}\label{ap-con-prob-pm-generic-comoving}
P^{\pm}_{\phi\rightarrow\gamma^{\pm}} (l)=  \frac{|A_{\pm} (l)|^2}{|\phi (l_0)|^2}  =   \frac{1}{4} g^2_{\phi\gamma} \int_{l_0}^{l} \frac{dl_1}{1+z} \int_{l_0}^{l} \frac{dl_2}{1+z} \left(\bar{B}_{\pm} (l_1) \, \bar{B}_{\pm}^* (l_2) \right) \, e^{-i\int_{l_2}^{l_1} \frac{dl_3}{1+z} \, \left[ \Delta_{\rm pl}(l_3) - \Delta_{\phi}(l_3) \right] } ~~,
\end{align}
%
Assuming the magnetic field to be homogeneous and redshifts as $\bar{B}_{\pm} = \bar{B}^0_{\pm}(1+z)^2$ where $\bar{B}^0_{\pm}$ is the comoving magnetic field at present epoch, and considering the expansion of the Universe, the conversion probability between redshifts $z \in [z_j, z_{j+1}]$ can be given as
\begin{align}\label{ap-con-prob-pm-z}
P^{\pm}_{\phi\rightarrow\gamma^{\pm}} (l) =   \frac{1}{4} g^2_{\phi\gamma}  \left(\bar{B}^0_{\pm} \, (\bar{B}^0_{\pm})^* \right) \, \int_{z_j}^{z_{j+1}} \frac{dz_1}{H}\int_{z_j}^{z_{j+1}} \frac{dz_2}{H}  \, (1+z)^2 \, e^{-i\int_{z_2}^{z_1} \frac{dz_3}{H(1+z)} \, \left[ \Delta_{\rm pl}(z_3) - \Delta_{\phi}(z_3) \right] } ~~,
\end{align}
where we used $dl = dz/H$, and $H(z)$ is the Hubble parameter at given redshift.
As we know at the resonance point, $\Delta_{\rm osc} (z_{\rm res}) = \Delta_{\rm pl}(z_{\rm res}) - \Delta_{\phi}(z_{\rm res}) =0$, therefore, phase factor integration must be done carefully. To resolve this issue, we can expand $\Delta_{\rm osc} (z)$ around the mid-point of each redshift interval, $z_c = \frac{z_j + z_{j+1}}{2}$. The appropriate discretization scheme is provided by the decreasing sequence of redshift, $z_{j+1} = z_j (1-\epsilon)$, where $\epsilon$ is the small parameter that sets the physical condition that the ALP to photon mixing propagation length remains smaller than the comoving Hubble radius $l \simeq \epsilon\, \mathcal{H}^{-1}$, in this work we set $\epsilon=0.1$~\cite{Addazi:2024mii}. Note that a smaller value of $\epsilon$ would increase the redshift bins and make the discretization smoother (at least around $z_{res}$). Furthermore, we can see that in the redshift interval where the resonance occur $z_c$ is related to $z_{res}$ by choosing $z_j = z_{\rm res} (1+\epsilon/2) $ and $z_{j+1} = z_{\rm res} (1-\epsilon/2) $, which gives $z_c=z_{\rm res}$. 
Following Ref.~\cite{Addazi:2024mii}, we use the semi-steady approximation to solve the integration, therefore in interval $z \in [z_j, z_{j+1}]$, we approximate $\Delta_{\rm osc} (z)$ around the mid-point as $\Delta_{\rm osc} (z) \simeq \Delta_{\rm osc} (z_c) + \Delta_{\rm osc}'(z_c) (z-z_c)$, where $\Delta_{\rm osc} (z) = \Delta_{\rm pl}(z) - \Delta_{\phi}(z)$ where prime denotes derivative with respect to $z$, we obtain
\begin{align}\label{ap-con-prob-pm-z-2}
P^{\pm}_{\phi\rightarrow\gamma^{\pm}} (z_j,z_{j+1}) 
 =   \frac{1}{4} g^2_{\phi\gamma}  \left(\bar{B}^0_{\pm} \, (\bar{B}^0_{\pm})^* \right) \, \frac{(1+z_c)^2}{H_c^2} \int_{z_j}^{z_{j+1}} dz_1\int_{z_j}^{z_{j+1}} dz_2 \, e^{-i\,\frac{(z_1-z_2)}{2 H_c (1+z_c)} \left[2 \Delta_{\rm osc} (z_c) +\Delta_{\rm osc}' (z_c) (z_1+z_2 - 2z_c) \right] }
\end{align}
where $H_c = H(z_c)$.
After performing the above double integration, we obtain 
\begin{align}\label{ap-con-prob-pm-z-fin}
P^{\pm}_{\phi\rightarrow\gamma^{\pm}} (z_j,z_{j+1}) = \frac{\pi}{8 \, |\Delta_{\rm osc}' (z_c) | } g^2_{\phi\gamma}  \left(\bar{B}^0_{\pm} \, (\bar{B}^0_{\pm})^* \right) \, \frac{(1+z_c)^3}{H_c} \Big| \, {\rm Erf}[(1+i)\alpha_j] - {\rm Erf}[(1+i)\alpha_{j+1}] \,\Big|^2 
\end{align}
where ${\rm Erf}$ is the error function and ${\rm Erf}[(1-i) = {\rm Erf}[(1+i)^*$ is the complex conjugate of each other, which suggests that the conversion probability derived above is indeed real and non-negative. The dimensionless parameters $\alpha$ are given as
\begin{align}\label{ap-alpha-def}
    \alpha_j = \frac{\Delta_{\rm osc} (z_c) + (z_j -z_c) \, |\Delta_{\rm osc}'(z_c)|}{2 \sqrt{H_c (1+z_c) \, |\Delta_{\rm osc}'(z_c)|}} ~, \quad \text{and} \quad  \alpha_{j+1} = \frac{\Delta_{\rm osc} (z_c) + (z_{j+1} -z_c) \, |\Delta_{\rm osc}'(z_c) | }{2 \sqrt{H_c (1+z_c) \, | \Delta_{\rm osc}'(z_c) | }} ~.
\end{align}
\begin{figure*}[t!]
%
\includegraphics[height=1.5in,width=7.4in]{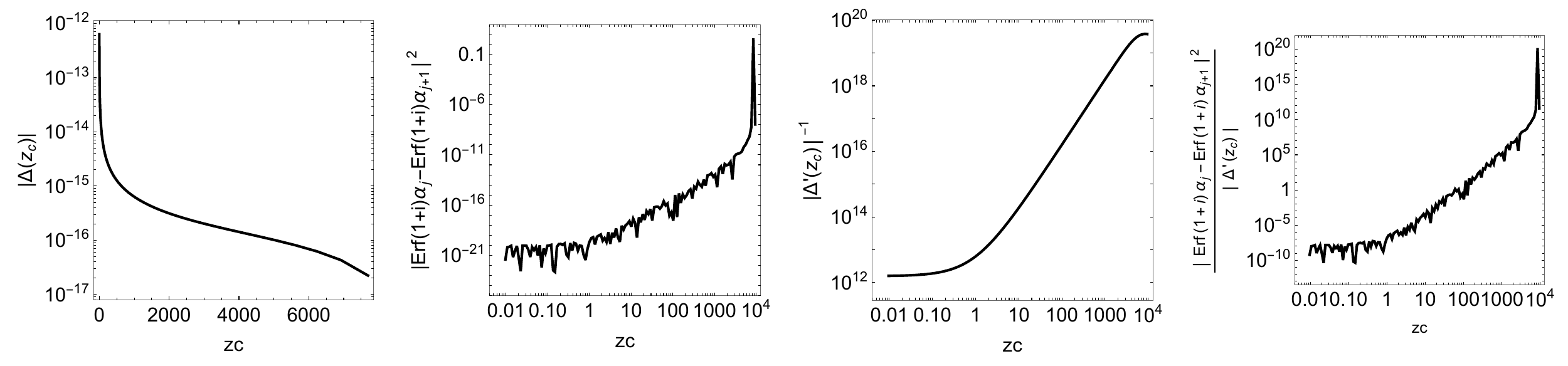} 
%
\includegraphics[height=1.5in,width=7.4in]{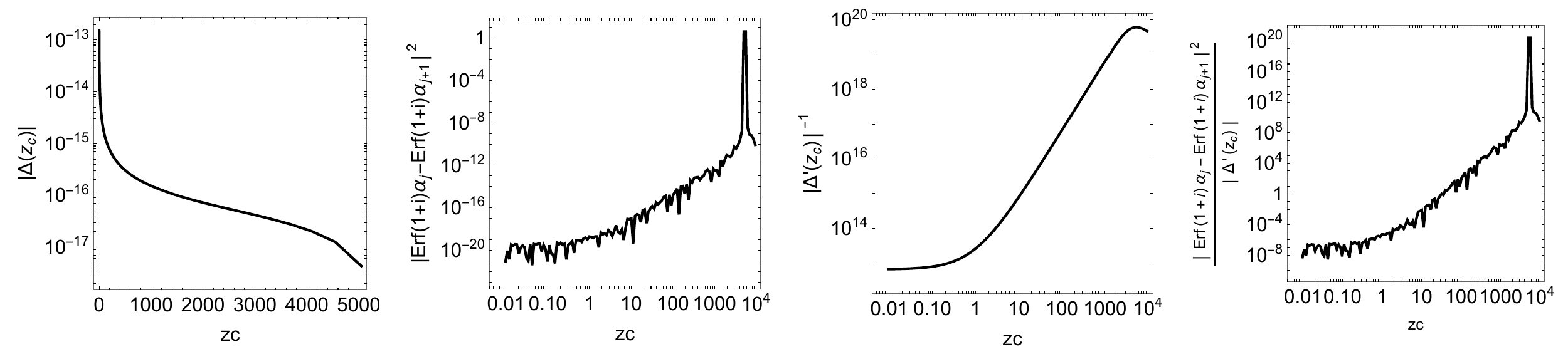} 
\includegraphics[height=1.5in,width=7.4in]{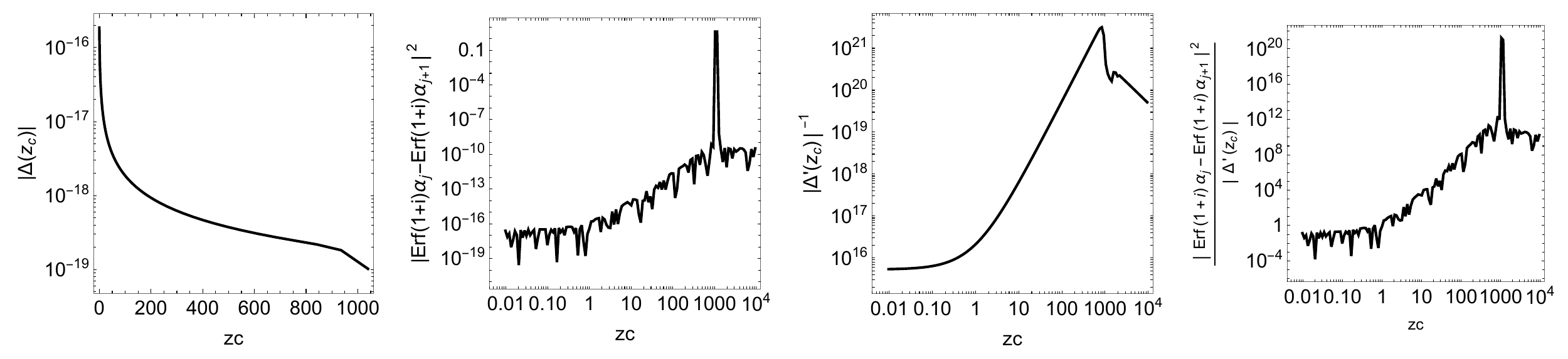} 
\caption{The behaviour of various terms in Eq.~\eqref{ap-con-prob-pm-z-fin-num} for three typical masses, $m_{\phi}=1.44301\times 10^{-8} \, {\rm eV} , 7.07209\times10^{-9} \, {\rm eV}  ,2.47505 \times 10^{-10}\, {\rm eV} $ going through the resonance at redshifts $z_c \simeq 8550,5314.41, 1094.18$ (upper, middle and lower panel), respectively.}
\label{appfig:prob-term}
\end{figure*}
Lastly, we give the conversion probability~\eqref{ap-con-prob-pm-z-fin} with typical numerical factors as
\begin{align}\label{ap-con-prob-pm-z-fin-num}
P^{\pm}_{\phi\rightarrow\gamma^{\pm}} (z_j,z_{j+1}) = \frac{\pi}{8 \, } \left( \frac{g_{\phi\gamma}}{{\rm GeV^{-1}}} \right)^2  \left( \frac{\left| \bar{B}^0_{\pm} \right|}{{\rm nG}} \right)^2\, \left( \frac{\rm eV}{4.96\times 10^{19}} \right)^2 \, \frac{(1+z_c)^3}{H_c\,|\Delta_{\rm osc}' (z_c) | } \left| \, {\rm Erf}[(1+i)\alpha_j] - {\rm Erf}[(1+i)\alpha_{j+1}] \,\right|^2 ~.
\end{align}
where we used $\bar{B}^0_{\pm} \, (\bar{B}^0_{\pm})^* =  \left| \bar{B}^0_{\pm} \right|^2$.
The above equation is an important relation that describes the probability of conversion of ALPs to chiral photons in the presence of a background helical magnetic field. We show in \ref{appfig:prob-term}, various terms in Eq.~\eqref{ap-con-prob-pm-z-fin-num} during the redshift interval $z_c\in [1000,10^4]$. As we can see, at the resonance frequencies, these terms show a peaked behavior, showing the resonance.

Before closing this section, we would like to mention an important point regarding the resonance.
For a given ALP mass $m_{\phi}$ which goes through a resonance at redshift $z_{res}=z_c$, the conversion probability in the corresponding redshift interval (around the resonance) with $\Delta (z_{res})=0$ has a simpler form as $\alpha_j = -\alpha_{j+1} = \frac{\epsilon \, z_{res} \,| \Delta_{\rm osc}'(z_{res}) | }{4 \sqrt{H_{res} (1+z_{res}) \, | \Delta_{\rm osc}'(z_{res}) | }}  = \alpha_{res}$. Therefore, we obtain
\begin{align}\label{ap-con-prob-pm-z-fin-num-res}
P^{\pm}_{\phi\rightarrow\gamma^{\pm}} (z_j,z_{j+1}) = \frac{\pi}{2 \, } \left( \frac{g_{\phi\gamma}}{{\rm GeV^{-1}}} \right)^2  \left( \frac{\left| \bar{B}^0_{\pm} \right|}{{\rm nG}} \right)^2\, \left( \frac{\rm eV}{4.96\times 10^{19}} \right)^2 \, \frac{(1+z_{res})^3}{H_{res}\,|\Delta_{\rm osc}' (z_{res}) | } \left| \, {\rm Erf}[(1+i)\alpha_{res}] \,\right|^2 ~,
\end{align}
which also depends on $\epsilon$. However, we have set $\epsilon=0.1$ for the reasons already discussed above.

\section{Properties of primordial helical magnetic field}\label{appsec-magnetic-field}
Let us briefly discuss the properties of the helical magnetic field, which are given by the magnetic field power spectrum~\cite{2013-Durrer.Neronov-Arxiv,2016-Subramanian-Arxiv} in Fourier space,
\begin{align}
    a^4 (\eta)\langle B_i (\textbf{k}) \, B_j^* (\textbf{q}) \rangle = (2\pi)^3 \delta^3 (\textbf{k}-\textbf{q})  \left[ (\delta_{ij} - \hat{k}_i \hat{k}_j) \, P_B (k)  - i\epsilon_{ijk} \hat{k}_m P_{aB} (k) \right]
\end{align}
where $a(\eta)$ is the scale factor in conformal time and the bracket $\langle ... \rangle$ denotes the ensemble average. $P_B$ and $P_{aB}$ are the symmetric and antisymmetric part of the magnetic power spectrum and are defined as
\begin{align}
    \langle B_+ (\textbf{k}) B_+^* (\textbf{q}) \rangle + \langle B_- (\textbf{k}) B_-^* (\textbf{q}) \rangle &= (2\pi)^3 \delta^3 (\textbf{k}-\textbf{q}) \, P_B (k)/a^4\\
    \langle B_+ (\textbf{k}) B_+^* (\textbf{q}) \rangle - \langle B_- (k) B_-^* (\textbf{q}) \rangle &= (2\pi)^3 \delta^3 (\textbf{k}-\textbf{q}) \, P_{aB} (k)/a^4 ~~.
\end{align}
%
Note that the above relations allows us to obtain $P_B\geq P_{aB}$, where equality holds where one component is vanishing and is referred to as maximally (or totally) helical magnetic field. Furthermore, $P_B$ and $P_{aB}$ are related to the magnetic energy density $\rho_B = \frac{1}{2} \langle \textbf{B(x)} \textbf{B(x)} \rangle$  and helicity density $\mathcal{H} = \textbf{A}\cdot\textbf{B}$ as
\begin{align}\label{rho-B-H}
    \rho_B = \frac{1}{a^4} \int \frac{dk}{k} \left( \frac{ k^3}{2\pi^2} P_B (k) \right), ~~
    \mathcal{H} = \int \frac{dk}{k} \left( \frac{ k^3}{2\pi^2} P_{aB} (k) \right) ~~.
\end{align}
In the Minkowski spacetime, $a(\eta) =1$, we obtain
\begin{align}
    \frac{d\mathcal{H}/d\ln{k}}{d\rho_B/d\ln{k}} = \frac{P_{aB} (k)}{P_{B} (k)} = \frac{\langle |B_{+}(k)|^2 \rangle - \langle |B_{-}(k) |^2 \rangle}{\langle |B_{+}(k)|^2 \rangle + \langle |B_{-} (k)|^2 \rangle} ~~.
\end{align}
Furthermore, similar to our previous works~\cite{2024-Kushwaha.Jain-PRD,Kushwaha:2025mia}, we can derive the following relation for the chirality parameter (equivalent to helicity fraction) for the helical magnetic fields
\begin{align}\label{chi-b}
     \frac{\langle |\bar{B}^0_{+}|^2 \rangle - \langle |\bar{B}^0_{-} |^2 \rangle}{\langle |\bar{B}^0_{+}|^2 \rangle + \langle |\bar{B}^0_{-} |^2 \rangle} = \frac{1-\delta_B}{1+\delta_B} = \Delta \chi_B
\end{align}
where $\delta_B = \langle | \bar{B}^0_{-}|^2 \rangle/\langle | \bar{B}^0_{+}|^2 \rangle = | \bar{B}^0_{-}|^2 / | \bar{B}^0_{+}|^2 $ defines the imbalance between the helicity modes of the the magnetic field, and $\Delta \chi_B$ is the chirality parameter that determines the net helicity or chirality of the magnetic field. As we can see that, $\Delta \chi_B=0$ for a non-helical magnetic field and $|\Delta \chi_B| = 1$ for a maximally helical magnetic field. The relation~\eqref{rho-B-H}, allows us to define a characteristic magnetic strength of the helical magnetic field as $\mathcal{B}_0 = \sqrt{2\rho_{B,0}} = \sqrt{ 2 (\langle |\bar{B}^0_{+}|^2 \rangle +  \langle |\bar{B}^0_{-} |^2 \rangle)}= \sqrt{ 2 (|\bar{B}^0_{+}|^2  + |\bar{B}^0_{-} |^2 )}$~\cite{2013-Durrer.Neronov-Arxiv}.
\begin{figure*}[t!]
%
\includegraphics[height=2.8in,width=3.4in]{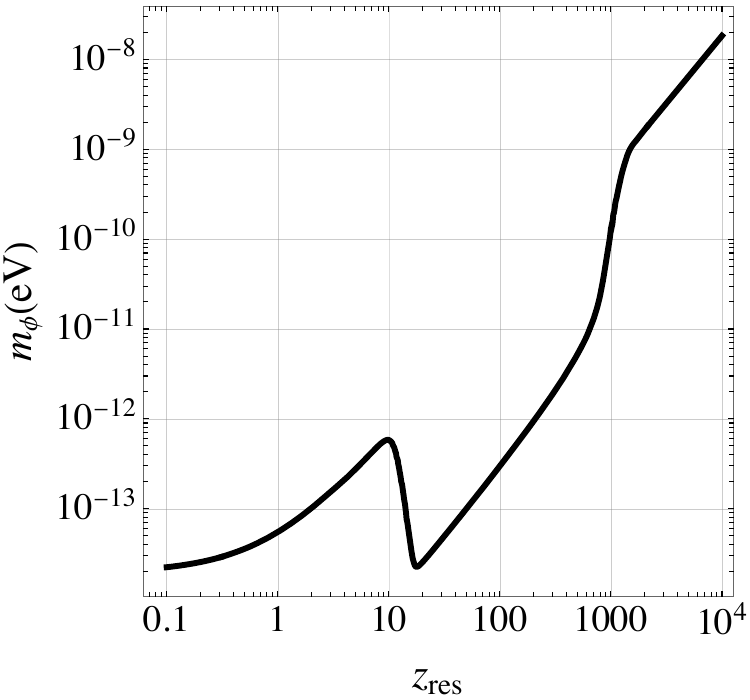} 
\hspace{.1cm}
\includegraphics[height=2.8in,width=3.3in]{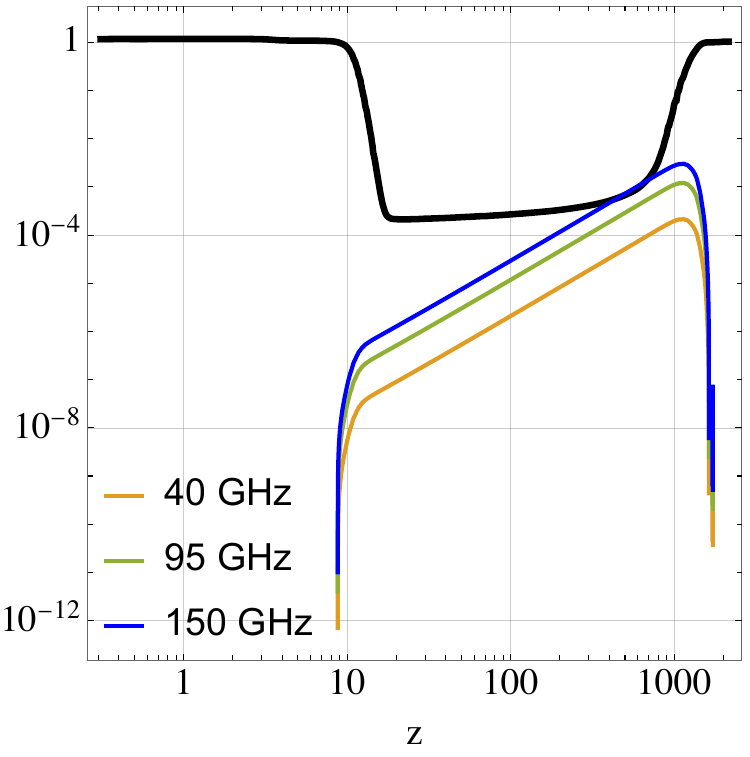} 
\caption{Left plot: shows the mass $m_{\phi}$ goes through resonance at redshifts $z_{res}$ in Eq.\eqref{res-mass-eq}. Right plot: shows the comparison of the plasma frequency term $X_e (z)$ (black curve) and frequency-dependent term, $7.3\times 10^{-3} \left(\frac{\omega_0 (z)}{1.5\times 10^{15} \, {\rm Hz}}\right)^2 (1+z)^2(1-X_e (z) \,)$ in Eq.\eqref{effective-mass-cmp} for three frequencies of our interest, we used $\omega (z) = \omega_0 (1+z)$ and $1\, {\rm eV} = 1.5\times 10^{15} \, {\rm Hz}$.}
\label{appfig:res-mass-term}
\end{figure*}

\section{Estimation of the resonant mass of ALPs}
\label{appsec-mass}
The resonant conversion of ALPs to photons occurs when the mass of ALPs, $m_{\phi}^2$ becomes equal to the effective mass of the photon, which in our case is determined by the plasma frequency $\omega_{\rm pl}^2$. Mathematically, this can be obtained by the condition, $\Delta_{\rm osc} (z_{res}) = \Delta_{\rm pl}(z_{res}) - \Delta_{\phi}(z_{res}) =0$. To see this explicitly, let us consider the following
\begin{align}\label{app-losc-eq}
    \Delta_{\rm osc} (z_{res}) &= \frac{\omega_{Pl,0}^2}{2\omega_0 \, (1+z_{res})} \left[ \left( \frac{m_{\phi}}{1.8\times 10^{-14} \, {\rm eV}} \right)^2 -   (1+z_{res})^3 X_e (z_{res}) \right] = 0 
\end{align}
which implies
\begin{align}\label{res-mass-eq}
    m_{\phi} = {1.8\times 10^{-14} \, {\rm eV}} \, (1+z_{res})^{3/2} X_e^{1/2} (z_{res}) ~,
\end{align}
therefore, we can obtain the resonant mass range as $m_{\phi} \in [2.58\times10^{-10}, 1.82\times 10^{-9}] \, {\rm eV}$ in $z_{res}\in [1100,2148]$. Furthermore, extrapolating $X_e (z)$ upto higher redshift with constant value, i.e, $X_e (z)= X_e (z=2148) $ for $2148 \leq z \leq 10^4$, gives the mass range $m_{\phi} \in [2.58\times10^{-10}, 1.82\times 10^{-8}] \, {\rm eV}$ in $z_{res}\in [1100,10^4]$. This extrapolation of $X_e(z)$ to $z=10^4$ is justified by following the Ref.\cite{Acharya:2023ygd}, see Ref.\cite{Cyr:2024sbd} for a recent development on modeling of the effective photon mass by the inclusion of Helium recombination.
Note that in Eq.\eqref{res-mass-eq}, we assume that the effective mass of the photon is determined only by the plasma frequency term, which gives the dominant contribution at the decoupling epoch.
However, often in the literature, the effective mass of the photon at a redshift is given by including the photon frequency-dependent term as (see Ref.~\cite{Mirizzi:2009nq,Mirizzi:2009iz,Kunze:2015noa}, for more details)
\begin{align}\label{effective-mass-cmp}
    \left( \frac{m_{\gamma}(z)}{{\rm eV}} \right)^2 \simeq 1.4\times10^{-21} \left[ X_e (z) - 7.3\times 10^{-3} \left(\frac{\omega (z)}{{\rm eV}}\right)^2 (1-X_e (z) \,) \right] \left( \frac{n_p (z)}{{\rm cm^{-3}}} \right)
\end{align}
 where $n_p$ is the proton number density. Because of the $\omega_0^2$ dependence the second term may dominate over the plasma frequency term (first term), and would lead to an imaginary effective mass $m_{\gamma}$. The effective mass squared has a positive contribution from the scattering off free electrons (i.e., first term or plasma frequency term) and a negative contribution from scattering off neutral atoms (i.e., second term or frequency-dependent term). Nevertheless, in our work, we focus only on the frequencies of CMB V-mode measurements, which are $\nu_0 = 33 \, {\rm GHz}, 40\, {\rm GHz}, 95\, {\rm GHz}, 150\, {\rm GHz}$. We should first verify whether the plasma frequency actually makes the dominant contribution during the epoch of our interest, i.e., $z\in [1100,10^4]$. To see this, we can only compare the terms in the square bracket in Eq.~\eqref{effective-mass-cmp}. 
%
As we can see from the \ref{appfig:res-mass-term} (right plot), the second term is only relevant for higher frequencies, $\nu_0=150 \, {\rm GHz}$ at redshift $300\lesssim z_{res} \lesssim 800$. Therefore, we can safely ignore the second term in our calculations and consider the effective mass to be dominated by the plasma frequency term only.

Before closing this section, we can define an important quantity 'the proper oscillation length' using Eq.~\eqref{app-losc-eq}, i.e., $2\ell_{\rm osc}^{-1} (z) =\Delta_{\rm osc} (z) = \Delta_{\rm pl} (z) - \Delta_{\phi} (z) = (m_{\phi}^2 - \omega_{\rm pl}^2)/(2\omega)$. Assuming the redshifting of frequency and constant $m_{\phi}$, we obtain
\begin{align}\label{losc-eq}
    \frac{\ell_{\rm osc}}{{\rm pc} } = \left(\frac{1+z}{1.93} \right) \left(\frac{\omega_0}{{\rm GHz}} \right) \, \left[ \left( \frac{m_{\phi}}{1.8\times 10^{-14} \, {\rm eV}} \right)^2 -   (1+z)^3 X_e (z) \right]^{-1}
\end{align}
where we used $\omega_{Pl,0}\simeq  1.8\times 10^{-14} \, {\rm eV}$ inside the bracket. Note that the comoving oscillation length can be obtained by $l_{\rm osc} = \ell_{\rm osc} \, (1+z)$ from the above equation.

\input{PRD-final.bbl}
\end{document}

%% file: PRD-final.bbl
%